# Graphene field-effect transistors with gigahertz-frequency power gain on flexible substrates

*Nicholas Petrone[†], Inanc Meric[‡], James Hone[†], and Kenneth L. Shepard\*[‡]*

[†]Department of Mechanical Engineering, [‡]Department of Electrical Engineering,

Columbia University, New York, New York 10027, United States

ABSTRACT

The development of flexible electronics operating at radio-frequencies (RF) requires materials that combine excellent electronic performance and the ability to withstand high levels of strain. In this work, we fabricate graphene field-effect transistors (GFETs) on flexible substrates from graphene grown by chemical vapor deposition (CVD). Our devices demonstrate unity-current-gain frequencies, $f_T$, and unity-power-gain frequencies, $f_{max}$, up to 10.7 and 3.7 GHz, respectively, with strain limits of 1.75%. These devices represent the only reported technology to achieve gigahertz-frequency power gain at strain levels above 0.5%. As such, they demonstrate the potential of CVD graphene to enable a broad range of flexible electronic technologies which require both high-flexibility and RF operation.

The field of flexible electronics has been active for more than 15 years, driven by the desire for low-cost, large-area, pliable electronics for such applications as e-paper, flexible displays, chemical and biological sensors, and smart tags.[1] The electronic materials used in these cases have largely been polymers and small-molecule organic films because of the desire to exploit large-area, low-cost fabrication approaches, such as roll-to-roll dry[2] or inkjet printing.[3] The resulting electronic device performance, however, has been relatively poor, with inherent low-field mobilities typically less than 1 $cm^2 V^{-1} s^{-1}$ and mobilities in integrated devices typically below 0.05 $cm^2 V^{-1} s^{-1}$.[1-3] Both reliability and low-voltage operation have been challenging. In addition, it is important for any proposed flexible technology to maintain uniform electronic properties over a wide range of strain, $\varepsilon$, which is related to the thickness, $t$, and bending radius, $\rho$, of the substrate as $\varepsilon = t/(2\rho)$.

The desire to improve the performance of these devices has led to growing efforts to transfer wires, ribbons, and membranes of traditional semiconducting materials to flexible substrates. Materials such as silicon nanomembranes (SiNMs),[4-7] III-V metal-oxide-semiconductor thin-films[8] and nanowires,[9] indium-gallium-zinc-oxide,[10] and AlGaN/GaN heterostructures[11] have been investigated, as have carbon nanotubes (CNTs),[12-15] and graphene.[16-17] However, enhancements to electronic performance have been achieved at the expense of device flexibility; to date, no flexible technology has achieved both unity-current-gain frequencies, $f_T$, and unity-power-gain frequencies, $f_{max}$, in the GHz regime at strains above 0.5%. CNT devices have demonstrated $f_T$ performance approaching 1 GHz at 1% strain for 0.8-μm channel lengths.[12-13] However, $f_{max}$, which is far more important for circuit applications, is not reported. In fact, $f_{max}$ is expected to be substantially less than $f_T$, following similar trends for field-effect transistors (FETs) based on mats of CNTs on rigid substrates.[18] The highest values of $f_{max}$ for FETs fabricated on flexible substrates have been reported for SiNMs[5] and III-V metal-oxide-semiconductor thin films[8] at 12 and 23 GHz, respectively. However, FETs based on these bulk semiconductor materials all exhibit strain limits below 0.5%.[4-5, 8, 11]

Graphene's unique electronic[19-20] and mechanical[21] properties make it a promising material for the fabrication of FETs which require both high flexibility and high operating frequencies. While graphene has no band-gap, rendering it poorly suited for digital applications, its high carrier mobility,[19, 22] saturation velocity,[23-25] and current-carrying capacity[26-27] make it a promising candidate for high-frequency analog applications. Graphene-based FETs (GFETs) fabricated on rigid substrates have in fact demonstrated values of $f_{max}$ of up to 34 GHz at channel lengths of 600 nm.[28]

Methods for producing graphene films suitable for flexible electronics include dielectrophoretic deposition of solution-processed graphene[16] and large-area growth of graphene by chemical vapor deposition (CVD).[29] GFETs from solution-processed methods demonstrate $f_T$ performance of approximately 2.2 GHz at 170-nm channel length under strain up to 0.5%.[16] However, poor electrostatics in these devices result in non-saturating current-voltage ($I – V$) characteristics and $f_{max}$ values of only 550 MHz.[16] In contrast, CVD graphene films display excellent electronic properties comparable to those of exfoliated graphene.[30] Even on flexible substrates, GFETs fabricated from CVD graphene exhibit field-effect mobilities up to 4,900 cm$^2$V$^{-1}$s$^{-1}$ [31] and maintain stable DC electronic properties at high levels of strain.[32-37] In this work we demonstrate GFETs fabricated from CVD graphene with $f_{max}$ of 3.7 GHz (at channel lengths of 500 nm) and strain limits of 1.75%. Fig. 1 compares the work presented here with other flexible high-frequency technologies on the merits of $f_{max}$ and strain limits, showing that the GFETs fabricated in this work are the first transistors to attain power gain in the GHz regime at strains above 0.5%.

Fig. 2a shows a schematic of the GFETs fabricated in this work. A bottom-gated device structure is implemented, motivated by previous work demonstrating that bottom-gated fabrication of GFETs with a dielectric layer applied over the gate electrode yields higher performance than top-gated devices which attempt to grow a gate oxide on the graphene surface.[28, 38] GFETs are fabricated on 127 μm thick polyethylene naphthalate (PEN) substrates (DuPont Teijin Films). Two-fingered bottom-gates (1nm Ti/30nm Au-Pd alloy) are defined by electron beam lithography and lift-off. The contact pad region of the gate is further thickened by subsequent patterning and evaporation of Ti/Au (1nm/50nm). A 6-nm gate dielectric of $HfO_2$ is conformally grown by atomic layer deposition (ALD) at 150 °C yielding a dielectric constant of $\kappa \approx 13$.[39] Large, single-crystals of graphene are grown by chemical vapor deposition (CVD) and transferred over the gate using well-established procedures.[30] Graphene is patterned with a second lithography step and reactive ion etching in an oxygen plasma. The devices are completed by evaporating Ti/Pd/Au (1nm/15nm/50nm) source and drain electrodes to contact the graphene. Devices are left uncapped. In addition, the thermal limits of the polymer substrate (~180 °C) prevent high-temperature thermal annealing processes from being used to remove resist residue on the graphene channel. Fig. 2b shows a cross-sectional schematic of a completed device. GFETs are fabricated with a gate length of 500 nm, source-to-drain spacing of 900 nm, and an effective channel width of 30 μm (two 15-μm wide gates in parallel). Fig. 2c shows an optical micrograph of a GFET device fabricated on a PEN substrate.

Electronic device characteristics are measured under ambient conditions. Samples are strained during electronic measurements by applying uniaxial tensile strain in the y-direction (see Fig. 2c) under two-point bending conditions, as shown in Fig. 2d. The strain, $\varepsilon_{yy}$, is calculated from the bending geometry using elastica theory assuming frictionless end-supports.[40]

Flexible GFET DC performance in the linear transport region is shown in Figs. 3a-c, where device resistance, $R$, is displayed against gate-to-source voltage, $V_{gs}$, taken with fixed source-to-drain voltage ($V_{sd}$ = 10 mV) at increasing strain from $\varepsilon_{yy}$ = 0% to 1.75%. Low-bias field-effect mobility, $\mu_{FE}$, is calculated from $\mu_{FE} = (L_{ch} g_m)/(W_{ch} C_{tot} V_{sd})$, where $L_{ch}$ is the channel length, $W_{ch}$ is the channel width, $C_{tot}$ is the total effective gate capacitance per unit area, and $g_m$ is the small-signal transconductance, defined as $(\partial I_d / \partial V_{gs})|_{V_{sd}}$, where $I_d$ is the measured drain current. $C_{tot}$ is determined by the series combination of the electrostatic capacitance, $C_e$, and the quantum capacitance, $C_q$. For the devices presented in this work, $C_e \approx 1700$ nF cm$^{-2}$, based on a parallel plate model. $C_q$ is density dependent over the charge carrier density range pertinent to this work ($n = 0.5 - 10 \times 10^{12}$ cm$^{-2}$), but it can be approximated as the mean of $C_q$ values calculated over this carrier density range. This approach, shown to be valid for similar devices over an equivalent carrier density range,[41] yields a constant value of $C_q \approx 2000$ nF cm$^{-2}$. These values of $C_e$ and $C_q$ result in $C_{tot} \approx 919$ nF cm$^{-2}$. The source-to-gate current, $I_{sg}$, is measured to remain below 0.5 pA over the entire strain range during device characterization, indicating negligible leakage current through the dielectric even at high strain. $\mu_{FE}$ for our flexible GFET is ~1,500 cm$^2$ V$^{-1}$s$^{-1}$ (for $V_{gs}$ = -0.25 V, the gate bias that yields the maximum $g_m$ for this device). This mobility is comparable to similar devices fabricated from exfoliated graphene on silicon substrates,[38, 41] demonstrating the excellent electronic quality of the CVD graphene utilized in this work. Although mobility remains relatively constant with strain up to $\varepsilon_{yy}$ = 1.75%, the position of the Dirac point with respect to $V_{gs}$ is observed to shift with increasing strain. We attribute this shift to changes in device electrostatics, related to mobile trapped charges in the gate-oxide and graphene-oxide interfaces, as the substrate is flexed. The presence of trapped charges in the gate-oxide, at the graphene-oxide interface, or in resist residue on the graphene surface additionally accounts for the hysteresis in the position of the Dirac point with respect to $V_g$ (~0.5 V) observed in low-bias measurements. We note, however, that the presence of residual resist residue from lithographic processing does not significantly contribute to the contact resistance between the graphene channel and evaporated electrodes, as the total contact resistance for this device is less than 300 Ω-μm, in the range best contact resistances reported for GFET devices (200-1,000 Ω-μm).[41-43] The ungated regions of the graphene channel will, however, effectively increase the contact resistance of the device. Improvements to the device architecture which act to minimize the gate-to-source and gate-to-drain spacer regions, such as by utilizing a self-aligned fabrication scheme, can further reduce the effective channel resistance.

Figs. 3d-f show $I-V$ characteristics for the same representative device, with $I_d$ plotted as a function of $V_{sd}$ at values of $V_{gs}$ decreasing from 0.25 V to -1 V in 0.25 V steps. Device characteristics represent a unipolar p-channel device. Devices are only measured up to $V_{sd} = 0.5$ V due to the thermal limitations of the polymer substrate. Above $V_{sd} = 0.5$ V the substrate melts locally under the device channel, causing both the substrate and overlaying GFET to mechanically warp in structure. $I-V$ characteristics are plotted for increasing levels of strain ranging from 0% to 1.75%. Changes in $I_d$ with increasing strain are correlated to the observed shifts in the Dirac point in Figs. 2a-c. At $\varepsilon_{yy} = 0\%$, measured values of $g_m$ and output resistance, $r_o$, are 5.1 mS and 259 Ω, respectively, at a bias point of $V_{gs} = -0.25$ V and $V_{sd} = 0.5$ V. We observe a maximum current density of 0.28 mA/μm, consistent with values reported for devices fabricated from CVD graphene of similar structure at equivalent electric fields.[31, 37, 44]

Fig. 4 shows RF characteristics for this same GFET device characterized in Fig. 3. Both current-gain ($h_{21}$) and unilateral power gain ($U$) are extracted from S-parameters measured at $V_{sd} = 0.5$ V. $V_{gs}$ values are chosen to maximize device transconductance; these values change with strain due to the Dirac point voltage shifts observed in Figs. 3a-c. The device demonstrates extrinsic cut-off frequency values (without any de-embedding) of $f_T$ = 7.2 GHz and $f_{max}$ = 2.6 GHz at a bias point of $V_{gs}$ = -0.25 V at $\varepsilon_{yy}$ = 0%, as shown in Fig. 4a. At $\varepsilon_{yy}$ = 1.25%, $f_T$ = 10.7 GHz and $f_{max}$ = 3.7 GHz are observed at $V_{gs}$ = 0.4 V (Fig. 4b). The RF performance does not degrade from its unstrained values up to strains of $\varepsilon_{yy}$ = 1.75% (Fig. 4c). We note that previously mentioned restrictions on the range of applied $V_{sd}$, resulting from thermal constraints of the polymer substrate, prevent strong current saturation and ultimately limit $f_{max}$ performance.[28] In spite of these limitations, these devices yield comparable performance to GFETs fabricated on rigid substrates with a similar layout.[45]

Fig. 5 shows the evolution of relevant device parameters with strain. Fig. 5a-d plot DC characteristics as a function of strain. Both $g_m$ (Fig. 5a) and $r_o$ (Fig. 5b) exhibit low variance (less than ±25%) up to strains of $\varepsilon_{yy}$ = 1.1%. At strains greater than 1.1%, variations in $r_o$ increase up to ±40%, likely related to the observed shifts in device electrostatics (most notably the shift in the Dirac point) with flexure. Fig. 5c plots device gate capacitance extracted directly from measured scattering parameters, $C_g$, as a function of strain. We note that the value of $C_g$ extracted at $\varepsilon_{yy}$ = 0% is 941 nF cm$^{-2}$, which matches well the expected value for $C_{tot}$ of 919 nF cm$^{-2}$ as described before. The variation in $C_g$ with strain is likely attributed to aforementioned shifts in trapped charges. Accounting for observed variations in $C_g$ with strain, values of $\mu_{FE}$ at strains above $\varepsilon_{yy}$ = 0% are calculated utilizing $C_g$ values extracted at equivalent strain states. Mobility for the device remains uniform with device flexure (Fig. 5d), exhibiting less than ±30% variance across the entire measured strain range up to $\varepsilon_{yy}$ = 1.75%, in good agreement with previous theoretical calculations[46] and experimental observations,[47] demonstrating the stability of the intrinsic electronic properties of the CVD graphene channel. Figs. 5e,f plot cut-off frequency as a function of strain. In all cases, $V_{sd}$ is 0.5 V and $V_{gs}$ is chosen at each measured strain point to maximize the device transconductance; these values of $V_{gs}$ are also shown in Fig. 5e. Both $f_T$ and $f_{max}$ demonstrate low variance (less than ±20%) with strain up to $\varepsilon_{yy} = 1.1\%$, above which an increase in both $f_T$ and $f_{max}$ of up to 40% is observed. We note that both DC and RF performance of the device remain uniform up to strains of $\varepsilon_{yy} = 1.1\%$; improvements to device structure which reduce trapped charges can allow for improved uniformity of electronic properties at strains greater than $\varepsilon_{yy} = 1.1\%$. Above strains of $\varepsilon_{yy} = 1.75\%$, most devices begin to fail as a result of cracking of the gate electrode, corresponding to clear irreversible degradations in electronic characteristics.

In conclusion, we demonstrate flexible GFETs fabricated from CVD graphene which display extrinsic values of $f_T$ and $f_{max}$ up to 10.7 GHz and 3.7 GHz, respectively, with strain limits of 1.75%. This is the first example of a flexible technology exhibiting both gigahertz-frequency power gain and strain limits above 0.5%. As such, this work demonstrates the potential of CVD graphene as a material to enable a wide-range of highly-flexible electronic technologies requiring analog FETs operating in the gigahertz frequency range.

FIGURES

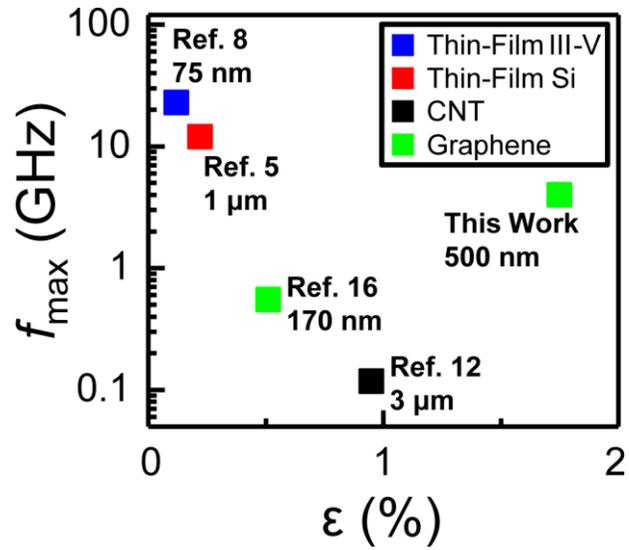

**Figure 1.** Comparison of $f_{max}$ and strain limits of flexible FET technologies. Channel lengths of the associated devices yielding these performances are noted.

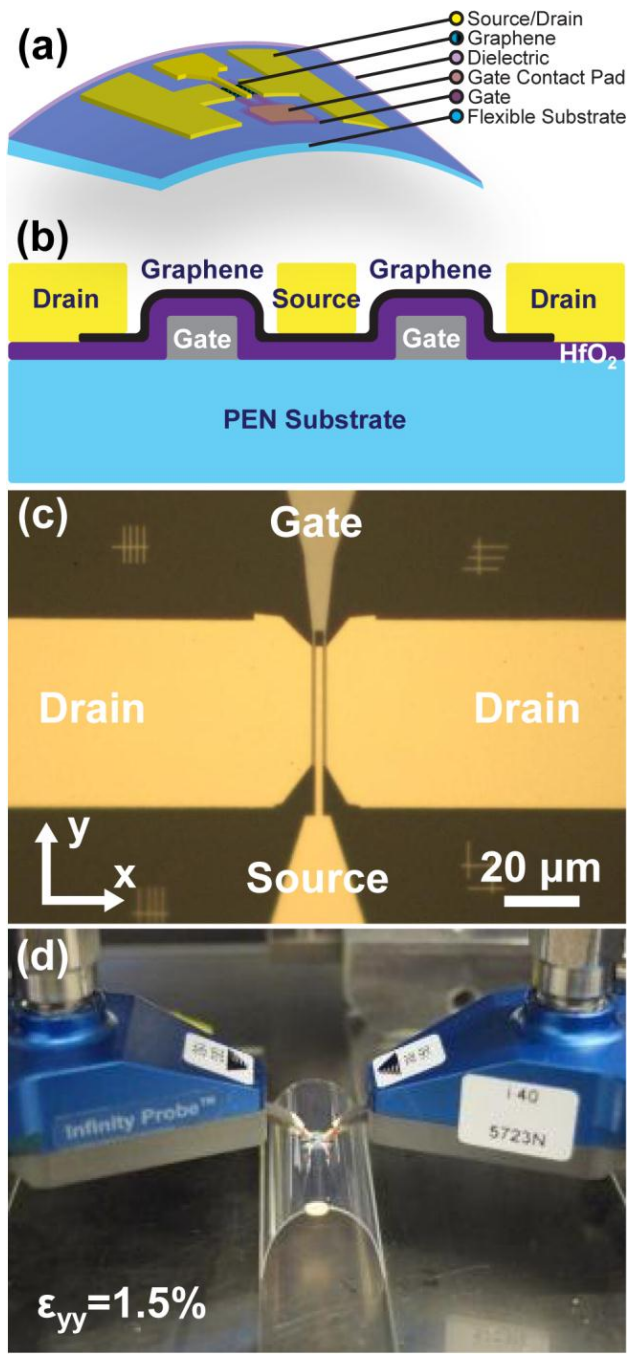

**Figure 2.** (a) Schematic of GFET fabricated on PEN, a flexible and transparent substrate. (b) Cross-sectional schematic of flexible GFET device. (c) Optical micrograph of GFET fabricated with a gate length of 500 nm and a source-to-drain spacing of 900 nm. (d) Photograph of electronic measurement approach for GFET under 1.5% strain.

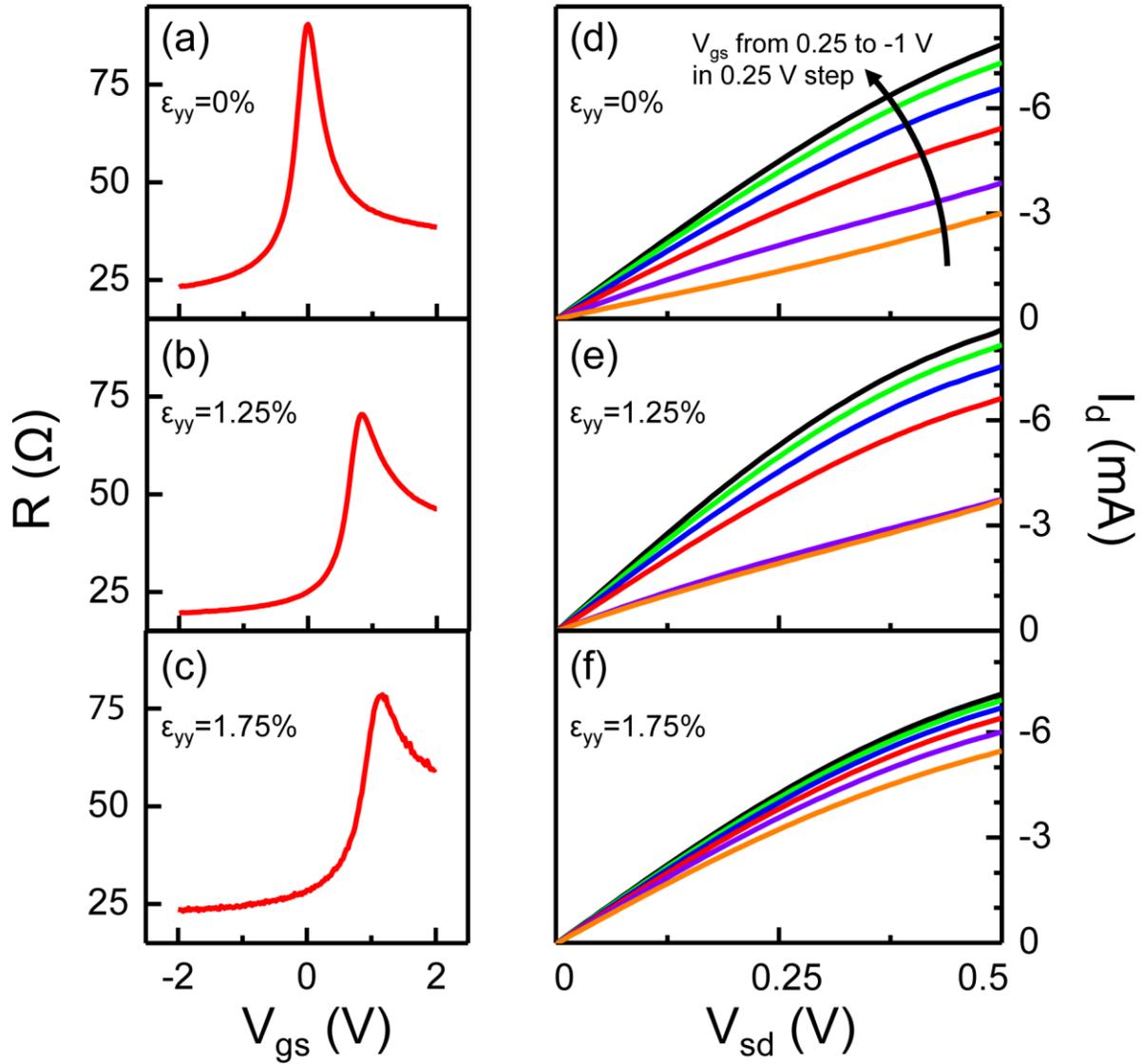

**Figure 3.** (a-c) Low-field transport characteristics of a flexible GFET with a device channel width of 30 μm. Device resistance, $R$, is plotted against gate-to-source voltage, $V_{gs}$, at a fixed source-to-drain bias of $V_{sd}$ = 10 mV. (d-f) Current-voltage ($I - V$) characteristics plotting drain current, $I_d$, as a function of $V_{sd}$. $I - V$ curves are taken at fixed $V_{gs}$ decreasing from 0.25 V (orange) to -1 V (black) in 0.25 V steps. Data are presented for increasing values of strain of $\varepsilon_{yy}$ = 0% (a,d), $\varepsilon_{yy}$ = 1.25% (b,e), and $\varepsilon_{yy}$ = 1.75% (c,f).

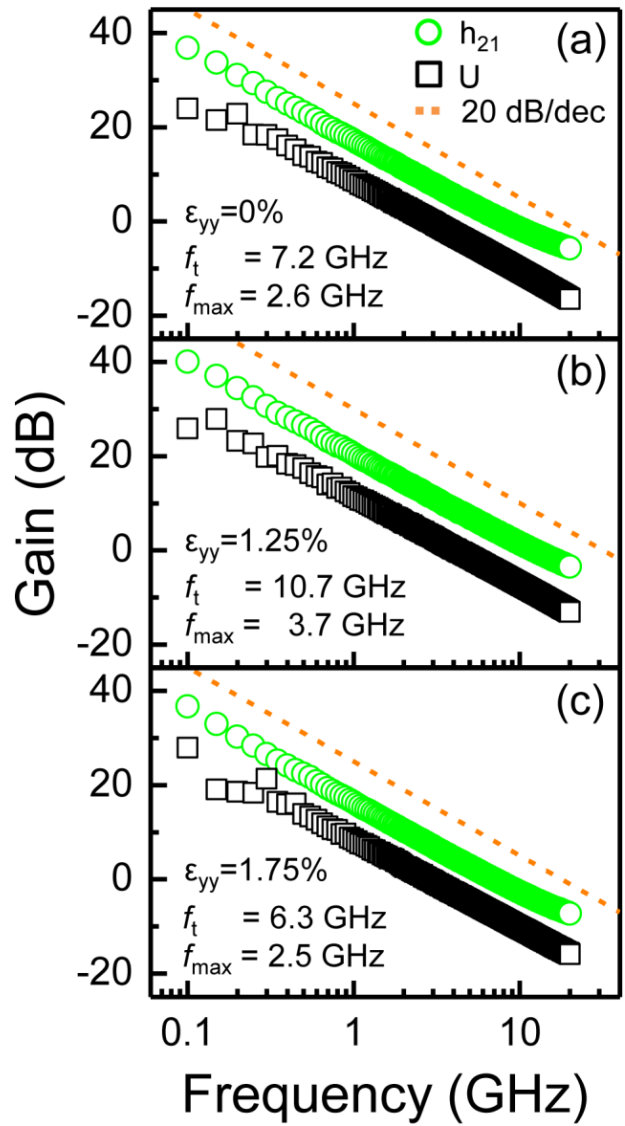

**Figure 4.** High-frequency device characteristics, current gain ($h_{21}$) and unilateral power gain ($U$), plotted as a function of frequency (without de-embedding). High-frequency characteristics are presented for strain values of $\varepsilon_{yy}$ = 0% (a), $\varepsilon_{yy}$ = 1.25% (b), and $\varepsilon_{yy}$ = 1.75% (c). Values of extrinsic $f_T$ and $f_{max}$ are calculated for each strain state. Measurements are performed at a fixed source-to-drain voltage, $V_{sd}$, of 0.5 V and gate-to-source voltages, $V_{gs}$, of -0.25 V (a), 0.4 V (b), and 0.6 V (c). The dashed line is a guide to the eye with a -20 dB/decade slope, included to demonstrate that devices follow well this expected frequency dependence.

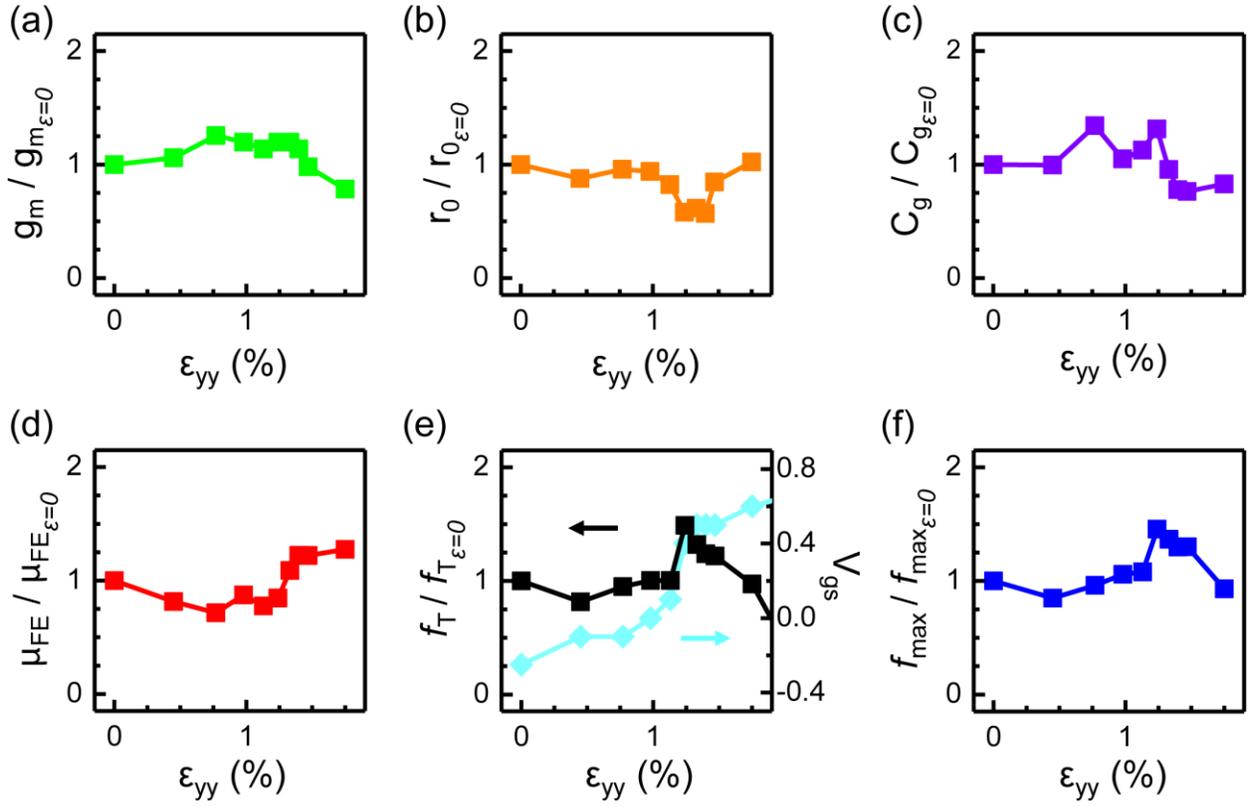

**Figure 5.** Device characteristics, normalized by their zero-strain values, plotted as a function of strain, $\varepsilon_{yy}$. Data is presented for (a) maximum transconductance, $g_m$; (b) maximum output resistance, $r_o$; (c) gate capacitance, $C_g$; (d) field-effect mobility, $\mu_{FE}$; (e) unity-current-gain frequency, $f_T$, (black) and the corresponding gate-to-source bias, $V_{gs}$, used to maximize transconductance (light blue); and (f) unity-power-gain frequency, $f_{max}$.

CORRESPONDING AUTHOR

*E-mail: shepard@ee.columbia.edu

FUNDING SOURCES

This work was funded by the Office of Naval Research under contract N00014-1210814, by the AFOSR MURI Program on new graphene materials technology, and by the Focus Center Research Program C2S2 Center.

ACKNOWLEDGMENT

N.P. thanks Prof. Jeffrey W. Kysar, Dr. Cory R. Dean, Dr. Gwan Hyoung Lee, and Dr. Arend M. van der Zande for helpful discussions.